\documentclass[fleqn,usenatbib,useAMS]{mnras}

\usepackage{graphicx}	
\usepackage{amsmath}	
\usepackage{multicol}        
\usepackage{bbm}		
\usepackage{pdflscape}	
\usepackage[switch,mathlines]{lineno} 
\usepackage{natbib}
\usepackage{mathrsfs}
\usepackage{comment}

\newcommand{\pd}{{\rm d}}

\newcommand{\id}{\mathbbm{I}}

\newcommand{\expt}{\mathbbm{E}}

\usepackage[T1]{fontenc}
\usepackage{ae,aecompl}

\usepackage{newtxtext,newtxmath}

\title[Rigorous galaxy correlation function]{Rigorous Formulation of Finite-Sample and Finite-Window Effects in Galaxy Clustering}

\author[T.\ T.\ Takeuchi et al.]{Tsutomu T.\ Takeuchi$^{1,2}$%
\thanks{E-mail: \href{mailto:tsutomu.takeuchi.ttt@gmail.com}{tsutomu.takeuchi.ttt@gmail.com} }, 
Satoshi Kuriki$^{3}$, 
and Keisuke Yano$^{4}$
\\
$^{1}$Division of Particle and Astrophysical Science, Nagoya University, Furo-cho, Chikusa-ku, Nagoya, 464–8602, Japan\\
$^{2}$The Research Center for Statistical Machine Learning, the Institute of Statistical Mathematics, 10-3 Midori-cho, Tachikawa, Tokyo 190–8562, Japan\\
$^{3}$School of Statistical Thinking, the Institute of Statistical Mathematics, 10-3 Midori-cho, Tachikawa, Tokyo 190–8562, Japan\\
$^{4}$Department of Fundamental Statistical Mathematics, the Institute of Statistical Mathematics, 10-3 Midori-cho, Tachikawa, Tokyo 190–8562, Japan
}

\date{}

\pubyear{{\the\year{}}}

\begin{document}
\label{firstpage}
\pagerange{\pageref{firstpage}--\pageref{lastpage}}
\maketitle

\begin{abstract}
Galaxy surveys provide finite catalogs of objects observed within bounded volumes, yet clustering statistics are often interpreted using theoretical frameworks developed for infinite point processes.
In this work, we formulate key statistical quantities directly for finite point processes and examine the structural consequences of finite-number and finite-window constraints.
We show that several well-known features of galaxy survey analysis arise naturally from finiteness alone.
In particular, non-vanishing higher-order connected correlations can occur even in statistically independent samples when the total number of points is fixed, and the integral constraint in two-point statistics appears as an exact identity implied by the finite-number condition rather than as an estimator artifact.
We further demonstrate that counts-in-cells and point-centered environmental measures correspond to distinct statistical ensembles.
Using Palm conditioning, we derive an exact relation between random-cell and point-centered statistics, showing that the latter probe a tilted version of the underlying distribution.
These results provide a probabilistic framework for separating structural effects imposed by finite sampling from correlations reflecting genuine astrophysical processes.
The formulation presented here remains valid for realistic survey geometries and finite data sets and clarifies the interpretation of commonly used clustering statistics in galaxy surveys.
\end{abstract}




\section{Introduction}

Two-point and higher-order correlation functions have long played a central role in the analysis of galaxy surveys, providing a quantitative description of spatial clustering over a wide range of physical scales \citep[e.g.,][]{1980lssu.book.....P}.
From the earliest measurements of galaxy clustering to modern large-area surveys, correlation functions have served as a bridge between observations and theoretical models of structure formation, encoding information about gravitational instability, biasing, and the growth of cosmic structure.

In practical analyses of galaxy surveys, non-zero higher-order (connected) correlations and the presence of the integral constraint are often interpreted as signatures of nonlinear clustering, biasing, or survey-specific systematics \citep{1977ApJ...217..385G, 1998ApJ...494L..41S}.
These features are typically viewed as reflecting genuine physical structure or, alternatively, as nuisances arising from finite survey geometry.

In most theoretical treatments, however, clustering statistics are formulated for \emph{infinite} point processes, often under assumptions of statistical homogeneity and isotropy \citep[e.g.,][]{1980lssu.book.....P}.
Under these assumptions, correlation functions admit elegant definitions and clear physical interpretations, and many standard results, such as the vanishing of higher-order \emph{connected} correlations for Poisson processes, follow immediately.

Observational data, by contrast, are necessarily finite: galaxy surveys cover limited volumes, contain finite numbers of objects, and are subject to selection effects and survey geometry.
This mismatch between theory and data is typically handled pragmatically.
Finite-volume effects are absorbed into empirical corrections, such as the integral constraint \citep[e.g.,][]{1977ApJ...217..385G, 1993ApJ...412...64L, 1999MNRAS.307..703R}, while non-zero higher-order correlations are interpreted as evidence for nonlinear gravitational evolution or complex biasing mechanisms \citep[e.g.,][]{1984ApJ...277L...5F, 1984ApJ...284L...9K, 1986ApJ...304...15B, 1998ApJ...494L..41S}.

Related issues have been discussed in the context of shot noise, finite-volume effects, and estimator bias.
However, these treatments typically regard finiteness as a secondary correction to an underlying infinite process.
Implicit in this approach is the assumption that the transition from an infinite point process to a finite sample is straightforward, and that finite samples can be regarded as faithful realizations of an underlying infinite process, up to small and correctable effects.

In this paper, we argue that this assumption is incomplete.
Starting from a formulation based on \emph{finite point processes}, we show that several well-known statistical features of galaxy survey data arise \emph{inevitably} from finite-number and finite-volume constraints alone, independent of any physical interactions.

Here, a \emph{finite point process} simply means a random \emph{finite} set of points observed in a bounded window, i.e., the mathematical object that directly represents a survey catalog.
This viewpoint emphasizes that the observed catalog itself is the primary probabilistic object, and that infinite-volume idealizations should be treated as limits rather than as implicit starting assumptions.

In particular, we demonstrate that
\begin{itemize}
\item non-vanishing higher-order (connected) correlations can arise even in completely independent samples when the total number of points is finite, and
\item the integral constraint in two-point statistics is an exact identity implied by the finite-number constraint,
\end{itemize}
so that these features are not anomalies or nuisances to be removed, but rather necessary consequences of the combinatorial structure of finite point sets.

This perspective has important implications for the physical interpretation of galaxy clustering measurements.
If finite-sample effects generate non-trivial correlations even in the absence of interactions, then observed deviations from naive Poisson expectations cannot automatically be attributed to astrophysical processes.
A careful separation between \emph{intrinsic} correlations, reflecting genuine physics, and \emph{induced} correlations, arising purely from finite sampling, is therefore essential.

Such a distinction is increasingly relevant for current and forthcoming surveys, where statistical precision and survey complexity both amplify the role of finite-volume structure.
Understanding which observed signatures arise from sampling constraints rather than from physical clustering is thus important for interpreting measurements of higher-order statistics, environmental trends, and bias.

Around the turn of the millennium, cosmological works have emphasized the impact of finite sampling on statistical inference in galaxy surveys.
In particular, studies by Colombi, Szapudi and collaborators \citep[e.g.,][]{1994A&A...281..301C, 1995ApJS...96..401C, 1996ApJ...465...14C, 1998MNRAS.296..253C, 1999MNRAS.310..428S, 2000MNRAS.313..711C, 2000MNRAS.313..725S} showed that finite catalogs naturally induce skewness, non-Gaussianity, and bias in measured clustering statistics, even when the underlying distribution is simple.

The present work shares with these studies the recognition that observational finiteness shapes statistical outcomes.
However, the conceptual standpoint differs in an essential way.
Earlier treatments primarily regarded finiteness as affecting the distribution of estimators around ensemble expectations.

Here, we instead emphasize that finiteness generates structural dependence in the point configuration itself.
From this viewpoint, features such as non-vanishing connected correlations or the integral constraint are not merely statistical fluctuations around an infinite-process limit, but intrinsic consequences of the fixed-number and finite-volume structure of the data.

Thus, rather than treating finite-sample effects as perturbations of an underlying infinite process, we regard the finite configuration as the primary probabilistic object.
In this sense, the present framework complements previous analyses by reinterpreting finite-sample effects as structural rather than purely inferential phenomena.

The aim of this work is to provide such a separation in a systematic and physically transparent manner.
By formulating correlation functions, factorial moments, and related quantities directly for finite point processes, we obtain expressions that remain exact at finite $N$ and finite volume.
Standard infinite-volume results then emerge as limiting cases, rather than as starting assumptions.

Throughout this paper, we focus on the physical interpretation and practical implications for galaxy surveys, keeping mathematical formalism to the minimum necessary for clarity.
More abstract questions concerning existence and infinite-volume limits of point processes are deferred to a companion paper aimed at a statistical audience.

This paper is organized as follows.
Section~\ref{sec:finite_point_process} introduces finite point processes as the natural probabilistic framework for galaxy survey data.
Section~\ref{sec:preliminaries} summarizes the basic statistical tools required for finite-sample analysis, including factorial moments, cumulants, and Palm conditioning.
Section~\ref{sec:finite_poisson} examines finite Poisson (binomial) processes and shows how non-vanishing connected correlations arise purely from finite-number constraints.
Section~\ref{sec:integral_constraint} revisits the integral constraint and demonstrates that it follows directly from the fixed total number of points.
Section~\ref{sec:cell_counts} analyzes the relation between random-cell and point-centered statistics using Palm expectations.
Finally, Section~\ref{sec:discussion} discusses the implications of these results for the interpretation of clustering measurements in galaxy surveys.
Technical details are summarized in the appendices.

\section{Finite Point Processes as the Natural Framework for Galaxy Surveys}
\label{sec:finite_point_process}

Galaxy surveys always consist of a \emph{finite catalog}:
a finite number of objects observed within a finite, and often complex, survey volume.
Regardless of the cosmological model or the assumed underlying matter distribution, the raw data delivered by an observation are a list of positions within a bounded region of space.
From this basic fact alone, the natural mathematical object describing survey data is a \emph{finite point process} \citep{2003Daley_point_processI, 2008Daley_point_processII, Baddeley2015}.

We therefore consider a bounded observation window $W \subset \mathbbm{R}^d$ and a finite point configuration $\mathbf X$ observed in $W$.
The precise notation and the basic probabilistic objects (counting measures, factorial moments, correlation functions, and Palm expectations) are summarized in Section~\ref{sec:preliminaries}.

In standard cosmological analyses, one often imagines that the observed catalog is a finite subsample drawn from an underlying infinite point process that satisfies statistical homogeneity and isotropy \citep{1980lssu.book.....P}.
While this picture is intuitively appealing and mathematically convenient, it is not operationally necessary.
All measurable quantities, such as counts in cells, correlation functions, and their estimators, are defined directly on the finite set $\mathbf X$.
No observable quantity directly accesses an underlying infinite realization; the infinite-volume limit enters only implicitly through assumptions about ensemble averages.

A finite point process formulation makes this situation explicit.
It treats the observed catalog as a complete probabilistic object in its own right, without assuming that it must be embedded in a larger, unobserved configuration \citep{Kallenberg2017}.
In this sense, finite point processes are not an approximation to the infinite case; rather, the infinite-point formalism should be understood as an idealized limit of the finite one.

This viewpoint is standard in spatial statistics, where point patterns are typically analyzed within bounded observation windows and finite-sample effects are treated as intrinsic features of the data \citep{Ripley1988, 2000StatisticaNeerlandica..54..329B}.
Adopting this perspective in the context of galaxy surveys provides a natural bridge between observational practice and probabilistic modeling, where finite-volume effects and internally estimated densities are unavoidable.

A key structural feature of finite point sets is that global constraints, such as the fixed total number of points, impose combinatorial relations among statistical quantities.
As a result, certain forms of correlation can arise even when point positions are independently sampled.
Effects such as non-vanishing higher-order correlations in Poisson samples, or the integral constraint in two-point statistics, emerge naturally from this finite structure \citep{Baddeley2015}.

Importantly, these effects are typically discussed in cosmology as finite-volume corrections or estimator-related systematics.
In the present formulation, however, they appear as intrinsic consequences of the probabilistic structure of finite catalogs rather than as secondary adjustments to an underlying infinite process.

From the finite point process perspective, these features are therefore not pathologies or corrections, but structural properties of finite samples.
Recognizing this distinction is useful for interpreting observed clustering signals, particularly when separating genuine physical correlations from those induced by sampling constraints.

The purpose of this section is to make this viewpoint explicit and intuitive, and to establish finite point processes as the natural starting point for the statistical analysis of galaxy surveys.
The formal statistical framework required to implement this perspective is introduced in the next section.
Subsequent sections then use it to identify which components of observed correlation functions carry genuine astrophysical information, and which arise unavoidably from finite-volume and finite-number effects.

\section{Preliminaries}
\label{sec:preliminaries}

In this section, we summarize the basic probabilistic objects needed to analyze finite point-process data.
Our goal is not to develop a full measure-theoretic theory of point processes, but to fix notation and record a minimal set of identities used throughout the paper.
The quantities introduced here correspond directly to commonly used observational statistics, such as counts-in-cells and correlation functions.

\subsection{Finite point processes and counting measures}

Let $W \subset \mathbbm{R}^d$ be a bounded observation window and let
\begin{align}
    \mathbf X = \{X_1,\dots,X_N\}, \qquad X_i \in W
\end{align}
denote a finite point configuration in $W$. 
Throughout this paper we restrict to $d=3$.
We represent the configuration by its counting measure
\begin{align}
    N(A) \equiv \sum_{i=1}^{N} \id\{X_i \in A\}, \qquad A \subset W,
\end{align}
where $\id$ denotes the indicator function \citep{2003Daley_point_processI, 2008Daley_point_processII, Kallenberg2017}.
Formally, the associated density field is written as
\begin{align}
    \rho(x) \equiv \sum_{i=1}^{N} \delta_{\mathrm D}(x-X_i),
\end{align}
where $\delta_{\mathrm D}$ denotes the Dirac delta function\footnote{Throughout this paper, spatial positions are denoted by $x$, $y$, \dots\ without boldface, for notational simplicity.}. 
This representation provides a convenient way to express familiar observational quantities within a finite-sample framework.

The finite-window mean density is defined by
\begin{align}
\bar{\rho} \equiv \frac{N}{|W|}.
\label{eq:mean_density_def}
\end{align}

\subsection{Factorial moments, cumulants, and correlation functions}

We characterize joint occurrence and connected structure using factorial moment densities $m^{[k]}$ and factorial cumulant densities $c^{[k]}$.
Correlation functions $\xi_k$ are defined by a normalization of $c^{[k]}$ with $\bar{\rho}$.
These quantities correspond to the hierarchy of clustering statistics used in galaxy surveys.
Precise definitions and the partition relations between $m^{[k]}$ and $c^{[k]}$ are summarized in Appendix~\ref{app:factorial} \citep{2003Daley_point_processI, 2008Daley_point_processII}.

\subsection{Palm distributions}

To describe statistics conditioned on the presence of a point at a given location, we introduce the reduced Palm expectation $\expt^{!x}$.
For a measurable functional $F$ of the point configuration, $\expt^{!x}[F]$ denotes expectation with respect to the point process conditioned on the existence of a point at $x$, with that point removed from the configuration \citep{palm1943, 2003Daley_point_processI, Kallenberg2017}.
A key identity is the Campbell--Mecke formula, recalled in Appendix~\ref{app:palm}.

Palm conditioning provides the natural framework for point-centered statistics and will be used in Section~\ref{sec:cell_counts} to relate environmental measures to counts-in-cells, which are widely employed in galaxy survey analyses.

\subsection{Finite-number constraint}

A central structural property of finite point sets is the global constraint that the total mass of $\rho$ in $W$ equals $N$.
Equivalently, for $\delta\rho(x)\equiv \rho(x)-\bar{\rho}$, one has
\begin{align}
    \int_W \delta\rho(x)\,\pd^3 x=0,
\end{align}
which expresses the fact that the mean density is internally determined from the sample itself.
This identity will play a central role in Section~\ref{sec:integral_constraint} (see Eq.~\ref{eq:finite_number_constraint}).

\section{Finite Poisson (Binomial) Processes: Basic Facts and Their Consequences}
\label{sec:finite_poisson}

As the most elementary example of a finite point process, we consider a \emph{binomial point process} defined in a bounded observation window $W \subset \mathbbm{R}^d$.
In this process, exactly $N$ points are present in $W$, and each point is distributed independently and uniformly over the window.
From a probabilistic point of view, this configuration represents a system without interactions and is often regarded as the canonical ``null model'' in spatial statistics and cosmology \citep{1980lssu.book.....P, Ripley1988, Baddeley2015}.

In spatial statistics, this model is understood as the finite-$N$ analogue of the homogeneous Poisson process \citep{2003Daley_point_processI, Kallenberg2017}.
In cosmological applications, Poisson processes are frequently used as the reference for ``no clustering'' \citep[e.g.,][]{1980lssu.book.....P}.

The key point emphasized in this section is that once the constraint $N<\infty$ is imposed, this point process no longer exhibits the property that higher-order connected correlations vanish.
Our aim is to demonstrate this fact explicitly using factorial moment densities and factorial cumulant densities \citep{2003Daley_point_processI, 2004ApJ...604...40T}, and to clarify how correlation functions should be interpreted for finite point-process data.

\subsection{Factorial moment densities and finite-number constraints}

For a binomial point process, the $k$-th order factorial moment density is independent of position and is given by
\begin{align}
m^{[k]}(x_1,\ldots,x_k)
=
\frac{N^{[k]}}{|W|^k},
\qquad
N^{[k]} \equiv \frac{N!}{(N-k)!}.
\label{eq:binomial_mk}
\end{align}
Although the point positions are independent and uniformly distributed, the factorial moment density is not simply $\bar{\rho}^k$, but explicitly contains the finite-number factor $N^{[k]}$. Thus, even at the level of factorial moments, the global constraint that exactly $N$ points are present is already encoded.

\subsection{Factorial cumulants and unavoidable connected structure}

Using the partition relations summarized in Appendix~\ref{app:factorial}, the corresponding factorial cumulant densities can be computed explicitly. Substituting Eq.~(\ref{eq:binomial_mk}) into the standard relations between factorial moments and factorial cumulants yields
\begin{align}
    c^{[k]}(x_1,\ldots,x_k)
    &= (-1)^{k-1}(k-1)! \frac{N}{|W|^k}, \notag \\
    &= (-1)^{k-1}(k-1)! \frac{\bar{\rho}}{|W|^{\,k-1}}, \qquad (k \ge 1).
\label{eq:binomial_ck}
\end{align}
The second line follows by introducing the mean number density $\bar{\rho} \equiv N/|W|$.
A crucial consequence of this result is that $c^{[k]}$ does \emph{not} vanish for $k \ge 2$. In other words, even for a point process that is completely independent in its construction, higher-order connected contributions appear once the total number of points is finite. These terms arise purely from the global constraint that exactly $N$ points must be present in the observation window.

These connected contributions do not originate from physical interactions or clustering.
They arise purely from the global finite-number constraint.
In finite point processes, connected structure therefore has an unavoidable combinatorial component that must be distinguished from genuine interaction effects \citep{Baddeley2015}.

In cosmological analyses, deviations from Poisson expectations are often interpreted in terms of nonlinear evolution or bias.
The result above shows that even the simplest independent sampling model can exhibit connected structure once finite-number constraints are taken into account.
This observation does not negate the physical origin of clustering, but indicates that the presence of non-zero higher-order correlations in finite samples does not by itself provide unambiguous evidence for interactions.

\subsection{Correlation functions for finite point processes}

Let $\bar{\rho}$ denote the finite-window mean density defined in Eq.~(\ref{eq:mean_density_def}).
The correlation functions $\xi_k$ are defined by normalizing $c^{[k]}$ with $\bar{\rho}$ as summarized in Appendix~\ref{app:factorial}.
For the binomial process, one obtains
\begin{align}
    \xi_k(x_1,\ldots,x_k)     &= (-1)^{k-1}(k-1)!\, \left(\bar{\rho} |W|\right)^{-(k-1)}, \notag \\
    &= (-1)^{k-1}(k-1)!\, N^{-(k-1)}, 
    \qquad (k \ge 2). \label{eq:binomial_xik}
\end{align}
Thus, in finite point processes, correlation functions quantify connected structure that includes both physical interactions and purely combinatorial constraints imposed by the finiteness of the sample.

\subsection{Infinite-volume limit}

Taking the infinite-volume limit
\begin{align}
|W| \to \infty,
\qquad
N \to \infty,
\qquad
\bar{\rho} = \mathrm{const.},
\label{eq:thermodynamic_limit}
\end{align}
one recovers
\begin{align}
c^{[k]} \to 0
\qquad (k \ge 2),
\label{eq:poisson_limit}
\end{align}
corresponding to the familiar property of an infinite Poisson process that higher-order connected correlations vanish.
From the present perspective, this behavior should be understood as a limiting case of finite processes rather than a property that can be assumed \emph{a priori} for finite observational data sets.

\subsection{Interpretation}

The binomial point process provides the clearest example of the conceptual difference between finite and infinite point processes.
Even in the absence of any interaction, finite point sets necessarily exhibit non-zero higher-order correlations due solely to global constraints.
This implies that detecting non-zero correlation functions in finite data does not automatically indicate physical clustering.
A careful separation between combinatorial and physical contributions is essential, particularly when interpreting higher-order statistics derived from finite survey volumes.
Detailed derivations are given in Appendix~\ref{app:factorial}.

\section{The Integral Constraint Revisited}
\label{sec:integral_constraint}

A well-known feature of galaxy survey analysis is the so-called \emph{integral constraint}, which enforces that the estimated density contrast integrates to zero over the survey volume \citep{1977ApJ...217..385G}.
In the standard cosmological literature, this constraint is often introduced as a correction associated with finite survey geometry or with the use of an internally estimated mean density \citep{1993ApJ...417...19H, 1993ApJ...412...64L}.
In the finite point process framework adopted here, however, the integral constraint arises in a more fundamental way: it is a direct consequence of the fixed total number of points in the observation window.

Let $W \subset \mathbbm{R}^d$ be a finite observation window containing exactly $N$ points, and let $\bar{\rho}$ be the finite-window mean density defined in Eq.~(\ref{eq:mean_density_def}).
Define the fluctuation field
\begin{align}
\delta\rho(x) \equiv \rho(x) - \bar{\rho}.
\end{align}
Then the finite-number constraint implies the identity
\begin{align}
    \int_W \delta\rho(x)\,\pd^3 x = 0.
\label{eq:finite_number_constraint}
\end{align}
This relation follows directly from the definition of $\bar{\rho}$ and holds independently of any assumptions about homogeneity or clustering.

Taking the second moment of this identity yields
\begin{align}
0
=
\int_W \int_W
\langle \delta\rho(x)\,\delta\rho(y) \rangle
\,\pd^3 x\,\pd^3 y .
\label{eq:delta_rho_identity}
\end{align}
Using the definition of the two-point correlation function for a finite point process,
\begin{align}
\langle \delta\rho(x)\,\delta\rho(y) \rangle
=
\bar{\rho}^2 \xi_2(x,y)
+
\bar{\rho}\,\delta_{\mathrm{D}}(x-y),
\end{align}
and integrating over the window, the delta-function term gives
\begin{align}
\int_W \int_W \delta_{\mathrm{D}}(x-y)\,\pd^3 x\,\pd^3 y = |W|.
\end{align}
One therefore obtains the exact identity
\begin{align}
\int_W \int_W \xi_2(x,y)\,\pd^3 x\,\pd^3 y
=
-\frac{|W|}{\bar{\rho}}.
\label{eq:integral_constraint_finite}
\end{align}
This relation holds for any finite point process, including the binomial process discussed in Section~\ref{sec:finite_poisson}.
It reflects the algebraic structure imposed by the finite-number constraint rather than any physical interaction or clustering mechanism.

In contrast, traditional treatments often assume an infinite point process with an externally specified mean density.
Under this assumption, the condition
\begin{align}
\int \xi_2(r)\,\pd^3 x = 0
\end{align}
is imposed \citep{1980lssu.book.....P}, and deviations from this relation in finite surveys are interpreted as an ``integral constraint'' correction \citep{1977ApJ...217..385G, 1993ApJ...417...19H, 1993ApJ...412...64L}.

From the present viewpoint, such corrections arise from applying an infinite-volume identity to data that are intrinsically finite.
The finite point process formulation clarifies that the integral constraint is not an ad hoc adjustment but an intrinsic property of finite samples.

For galaxy surveys, this interpretation shifts the role of the integral constraint from that of an estimator correction to that of a structural identity imposed by finite catalogs.
This viewpoint helps separate features induced by survey finiteness from those reflecting genuine large-scale structure.

This perspective parallels spatial statistical treatments, where finite-window effects are regarded as structural consequences of bounded observation domains rather than estimator deficiencies \citep{Ripley1988, Baddeley2015}.

The result provides a natural conceptual basis for understanding finite-window effects and prepares the ground for the conditional statistics introduced in the next section, where Palm conditioning is used to describe point-centered averages in a consistent manner.

\section{Counts-in-Cells and Point-Centered Statistics}
\label{sec:cell_counts}

Counts-in-cells statistics are widely used to quantify spatial structure in point distributions \citep{1980lssu.book.....P, 1998ApJ...494L..41S}.
In practice, however, two conceptually distinct constructions are employed.

One may either place cells randomly in space and count the number of points contained within them, or one may center cells on observed points and count their neighbors.
Although these procedures are often treated interchangeably in applications, they correspond to fundamentally different averaging operations.

Within the finite point process framework, this distinction can be made precise.
Let $C$ denote a reference cell and define its translated version by $C_x \equiv x + C$.
The number of points contained in the windowed cell is
\begin{align}
N_x \equiv N(W \cap C_x).
\end{align}

Two natural ensemble averages arise:
\begin{itemize}
\item[(R)] \textbf{Random-cell averaging:} the cell center $x$ is chosen independently of the point process.
\item[(P)] \textbf{Point-centered averaging:} the cell is centered on a point $x \in \mathbf X$.
\end{itemize}
While both constructions probe local environments, they represent distinct statistical viewpoints.
The first samples space uniformly, whereas the second samples \emph{points} uniformly.
This distinction becomes particularly significant in finite samples.

In point process theory, the point-centered construction is formalized through Palm distributions, which describe the conditional structure of a process given that a point is present at a specified location \citep{palm1943, 2003Daley_point_processI, Kallenberg2017}.

\subsection{Generating functions}

Define the probability generating function for the random-cell count by
\begin{align}
G_R(z;x)
\equiv
\mathbbm{E}\!\left[z^{N_x}\right].
\end{align}

For the point-centered count, define the neighbor count
\begin{align}
N_x^{\mathrm{neigh}}
\equiv
N\!\left((W \cap C_x)\setminus\{x\}\right),
\end{align}
and the corresponding generating function
\begin{align}
G_P(z;x)
\equiv
\mathbbm{E}^{!x}
\!\left[
z^{N_x^{\mathrm{neigh}}}
\right],
\end{align}
where $\mathbbm{E}^{!x}$ denotes the reduced Palm expectation at location $x$.

\subsection{Exact relation between random-cell and point-centered statistics}

Applying the Campbell--Mecke identity (Appendix~\ref{app:palm}) with
$F(x,X)=z^{N_x^{\mathrm{neigh}}}$ yields the exact relation
\begin{align}
\mathbbm{E}[N_x]\,
G_P(z;x)
=
\mathbbm{E}
\!\left[
N_x
\,z^{\,N_x-1}
\right].
\label{eq:palm_count_identity}
\end{align}
Since
\begin{align}
\mathbbm{E}
\!\left[
N_x
\,z^{\,N_x-1}
\right]
=
\frac{\partial}{\partial z}
G_R(z;x),
\end{align}
it follows that
\begin{align}
G_P(z;x)
=
\frac{1}{\mathbbm{E}[N_x]}
\,
\frac{\partial}{\partial z}
G_R(z;x).
\label{eq:palm_tilt_exact}
\end{align}
Equation~(\ref{eq:palm_tilt_exact}) shows that point-centered statistics are obtained from random-cell statistics through an \emph{exact} size-bias (tilt) transformation, with a normalization that does not require homogeneity, stationarity, or an infinite-volume limit.

\subsection{Interpretational implications}

Equation~(\ref{eq:palm_tilt_exact}) clarifies that point-centered environmental measures are not merely alternative implementations of counts-in-cells but correspond to a different statistical ensemble.
Palm conditioning reweights configurations toward regions of higher occupancy, reflecting the fact that points are more likely to reside in denser environments.
Thus, nearest-neighbor measures and environment indicators probe a size-biased version of the underlying distribution.

For galaxy surveys, this implies that point-centered measures do not sample the same statistical ensemble as randomly placed cells.
Instead, they systematically emphasize higher-density environments.
Recognizing this distinction helps clarify the interpretation of environment-dependent trends in galaxy properties, which may partly reflect statistical conditioning rather than purely physical segregation.
Such considerations may be particularly relevant when interpreting environment-dependent trends in galaxy properties \citep[e.g.,][]{1980ApJ...236..351D, 1997ApJ...490..577D, 2005ApJ...629..143B, 2006MNRAS.370..198C}.

Further, we can also connect the size-bias relation to the statistics related to galaxy formation, such as bias or peak formalism. 
In the language of density fields, the average over galaxy positions can be written as
\begin{align}
\langle A\rangle_{\rm galaxy}
=
\frac{\langle n(x)A\rangle}{\langle n(x)\rangle},
\end{align}
which represents a density-weighted average. 
This structure appears widely in cosmological contexts such as peak bias and environment-dependent statistics.
The Palm conditioning relation
\begin{align}
\mathbbm{E}^{!x}[A]
=
\frac{1}{\rho(x)}
\mathbbm{E}
\left[
n(x)A
\right]
\end{align}
coincides exactly with this density-weighted average.
Therefore, the size-bias relation can be regarded as the probabilistic formulation of galaxy-position averages and peak-bias statistics commonly used in cosmology.

Here, we should note that this Palm-induced distinction is independent of the finite-number constraint. 
While constraint-induced correlations vanish in the infinite-volume limit, the difference between point-centered and random sampling persists whenever intrinsic correlations are present, even in statistically homogeneous systems. 

\section{Discussion and Conclusion}
\label{sec:discussion}

The results presented in this work suggest a shift in perspective on several standard statistical constructs used in the analysis of galaxy surveys.

Rather than viewing effects such as higher-order correlations, the integral constraint, or differences between point-centered and randomly placed environment measures as methodological complications, the finite point process framework shows that these features arise naturally from the structure of finite data.
They are therefore not anomalies to be corrected but intrinsic properties of the sampling process itself.

This perspective has two important implications.
First, it clarifies that the presence of non-zero higher-order correlation functions does not necessarily indicate nonlinear clustering or interactions.
Finite-number constraints alone can generate connected structure even in statistically independent point configurations.
As a result, the interpretation of higher-order statistics requires a careful separation between finite-sample effects and genuine astrophysical signals.
This distinction is particularly relevant in light of the traditional use of Poisson sampling as a null hypothesis for the absence of clustering \citep{1980lssu.book.....P}.

Second, it provides a unified interpretation of commonly used environmental measures.
Point-centered statistics are shown to probe a Palm-conditioned (tilted) version of the underlying distribution, systematically emphasizing denser regions.
This observation connects nearest-neighbor statistics, environment indicators, and counts-in-cells within a single probabilistic framework and highlights that their differences are structural rather than purely operational.

From a cosmological standpoint, these insights are particularly relevant for current and forthcoming surveys, where finite survey volumes, complex masks, and spatially varying selection functions are unavoidable.
Interpreting clustering measurements without explicitly accounting for the finite nature of the data risks conflating sampling structure with physical information.
In this sense, part of the signal traditionally attributed to clustering strength or environmental dependence may reflect structural properties of finite sampling.
We emphasize that this Palm-type distinction is conceptually different from the finite-number constraint effects. The latter disappear in the infinite-volume limit, whereas the former arises from sampling at point locations and remains whenever genuine correlations exist, even in homogeneous systems.

The finite point process formulation therefore offers a conceptual basis for distinguishing between structural effects imposed by observation and signals arising from galaxy formation and large-scale structure.
Rather than replacing existing clustering analyses, it complements them by clarifying which statistical features are inevitable consequences of finiteness and which encode genuine astrophysical information.

Embedding familiar measures within a finite probabilistic framework thus provides a basis for interpreting observations in a way that remains valid for realistic survey geometries and finite data sets.
This perspective becomes increasingly important as survey precision improves and statistical uncertainties approach the level at which finite-sample structure is no longer negligible.
Finite galaxy catalogs should therefore not be regarded merely as approximations to infinite point processes, but as statistical systems with their own internal structure imposed by finite sampling.

Future work may extend this framework toward practical estimator construction and toward more explicit connections with cosmological modeling.
In particular, incorporating finite-sample structure into forward models of galaxy clustering may help clarify the relationship between observed statistics and the underlying physical processes.

\section*{Acknowledgements}
We thank Shiro Ikeda, Kenji Fukumizu, and Evgeny Spodarev for the fruitful discussion on the statistical aspect of the data analysis. 
We have been supported by JSPS Grants-in-Aid for Scientific Research (24H00247). 
This work has also been supported in part by the Collaboration Funding of the Institute of Statistical Mathematics ``Machine-Learning-Based Cosmogony: From Structure Formation to Galaxy Evolution''. 

\section*{Data Availability}
No new data were generated or analysed in this study.

\bibliographystyle{mnras}
\bibliography{k_point_correlation}

\appendix

\section{Factorial Moments and Cumulants}
\label{app:factorial}

This appendix summarizes the relations between factorial moments and factorial cumulants for finite point processes.
These relations provide the formal basis for the decomposition of correlation functions used in the main text and clarify how connected structure is separated from purely combinatorial contributions.

\subsection{Factorial moment hierarchy}

For a finite point process in a bounded window $W$, the factorial moment densities $m^{[k]}$ are defined (for distinct points) through
\begin{align}
\expt \left[
\prod_{j=1}^{k} N(\mathrm d x_j)
\right]
=
m^{[k]}(x_1,\dots,x_k)\,
\prod_{j=1}^{k}\mathrm d x_j .
\end{align}
For distinct points, these quantities describe the joint occurrence structure of $k$ points in infinitesimal neighborhoods.

\subsection{Cumulant expansion}

Connected structure is described by the factorial cumulant densities $c^{[k]}$.
They are related to factorial moments through the partition expansion \citep{2003Daley_point_processI}
\begin{align}
m^{[k]}(x_1,\dots,x_k)
=
\sum_{\pi \in \mathcal P_k}
\prod_{B \in \pi}
c^{[|B|]}(x_B),
\end{align}
where $\mathcal P_k$ denotes the set of partitions of $\{1,\dots,k\}$ and $x_B$ is the collection of arguments indexed by block $B$.

For example,
\begin{align}
m^{[2]}(x_1,x_2)
&=
c^{[2]}(x_1,x_2)
+
c^{[1]}(x_1)c^{[1]}(x_2),
\end{align}
\begin{align}
m^{[3]}(x_1,x_2,x_3)
&=
c^{[3]}(x_1,x_2,x_3) \notag \\
&\quad + \sum_{\mathrm{cyc}} c^{[2]}(x_i,x_j)c^{[1]}(x_k)
\notag\\
&\quad+ c^{[1]}(x_1)c^{[1]}(x_2)c^{[1]}(x_3).
\end{align}

\subsection{Correlation functions}

Let $\bar{\rho}$ be the finite-window mean density defined in Eq.~(\ref{eq:mean_density_def}).
We define correlation functions by the normalization
\begin{align}
    c^{[k]}(x_1,\dots,x_k) = \bar{\rho}^k \,\xi_k(x_1,\dots,x_k).
\end{align}

In finite samples, $\xi_k$ need not vanish even for independent sampling because $c^{[k]}$ can contain combinatorial contributions induced by the fixed-$N$ constraint.
This decomposition provides the basis for identifying which parts of observed correlations reflect intrinsic clustering and which arise structurally from finiteness.
These relations are used in Section~\ref{sec:finite_poisson}.

\section{Palm Conditioning and Size-Biased Statistics}
\label{app:palm}

This appendix outlines how point-centered statistics arise naturally from Palm conditioning and clarifies their exact relation to random-cell statistics in finite point processes.

\subsection{Palm expectation and Campbell--Mecke identity}

Let $F$ be a measurable functional of the point configuration.
The reduced Palm expectation $\expt^{!x}$ satisfies the Campbell--Mecke identity \citep{2003Daley_point_processI, Kallenberg2017}
\begin{align}
\expt
\left[
\sum_{x \in \mathbf X} F(x, \mathbf X\setminus\{x\})
\right]
=
\int_W
\bar{\rho}\,
\expt^{!x}[F(x,\mathbf X)]
\,\pd^3 x .
\label{eq:campbell_mecke}
\end{align}

\subsection{Cell counts and generating functions}

Let
\begin{align}
N_x \equiv N(W \cap C_x)
\end{align}
denote the count in a translated cell.

Random-cell statistics are described by
\begin{align}
G_R(z;x)
=
\expt[z^{N_x}],
\end{align}
while point-centered statistics involve the neighbor count
\begin{align}
N_x^{\mathrm{neigh}}
=
N\!\left((W \cap C_x)\setminus\{x\}\right)
\end{align}
and the generating function
\begin{align}
G_P(z;x)
=
\expt^{!x}
\!\left[
z^{N_x^{\mathrm{neigh}}}
\right].
\end{align}

\subsection{Exact tilt relation}

Applying Eq.~(\ref{eq:campbell_mecke}) with $F(x,\mathbf X)=z^{N_x^{\mathrm{neigh}}}$ yields
\begin{align}
\bar{\rho}\,G_P(z;x)
=
\dfrac{\displaystyle \expt \left[ \sum_{x\in \mathbf X} z^{N_x^{\mathrm{neigh}}} \right]}
{\displaystyle \int_W \pd^3 x}.
\end{align}

Equivalently, at the level of cell-centered statistics one obtains the exact identity
\begin{align}
\mathbbm{E}[N_x]\,
G_P(z;x)
=
\mathbbm{E}
\!\left[
N_x\,z^{N_x-1}
\right]
=
\frac{\partial}{\partial z}G_R(z;x),
\end{align}
and therefore the normalized tilt relation
\begin{align}
G_P(z;x)
=
\frac{1}{\mathbbm{E}[N_x]}
\,
\dfrac{\partial}{\partial z}
G_R(z;x).
\label{eq:app_palm_tilt_exact}
\end{align}
The normalization by 
\begin{align}
    \mathbbm{E}[N_x]=\dfrac{\partial}{\partial z} G_R(1;x)
\end{align}
ensures that the resulting law remains a proper probability distribution.
This relation holds for finite point processes without requiring assumptions of homogeneity or infinite volume.

In realistic surveys, masks, selection functions, and boundary truncations generally make $\mathbbm{E}[N_x]$ depend on the cell center $x$.
Equation~(\ref{eq:app_palm_tilt_exact}) nevertheless holds pointwise in $x$ as an identity of the finite point process, with the local normalization 
\begin{align}
    \mathbbm{E}[N_x]=\dfrac{\partial}{\partial z} G_R(1;x).
\end{align}
Thus, when comparing environments across the survey, one should distinguish genuine clustering from spatial variations induced purely by the survey window and selection effects.

\bsp	
\label{lastpage}
\end{document}